# Effect of different polarity solvents on total phenols and flavonoids content, and In-vitro antioxidant properties of flowers extract from *Aurea Helianthus*

Hyon-il Ri [1], Chol-song Kim[2,3], Un-hak Pak[1], Myong-su Kang[2], Tae-mun Kim[2*]

1. Kim Hyong Jik University of Education，Pyongyang 999093，Democratic People's Republic of Korea
2. Kim Il Sung University, Pyongyang 999095，Democratic People's Republic of Korea
3. Peking University, Beijing 100871, PR China

**Abstract**

The total phenols and flavonoids content of different polar solvent extracts from *Aurea Helianthus* flowers, and their antioxidant activity were determined. The ethanol extract of *Aurea Helianthus* flowers were suspended in water and fractionated using different polar solvents; hexane, chloroform, ethyl acetate, butanol and water. The parameters of each extract mentioned above were determined using Folin-Ciocalteu reagent (FCR) method, $AlCl_3$ colorimetry method, Ferric reducing ability of plasma (FRAP) assay, total antioxidant activity (TAA) assay and DPPH radical scavenging assay. The highest total phenols content (516.21 mg GAE/g) and flavonoids content (326.06 mg QCE /g）were obtained in ethyl acetate extract, the correlation between TPC and TFC assay was found to be 0.967. All polar solvent extracts of *Aurea Helianthus* flowers showed significant antioxidant effects, the highest inhibition was obtained in ethyl acetate and chloroform extracts and the lowest inhibition in the water extract. There is a good correlation of total phenols and flavonoids content with antioxidant activity. This work indicated that the polar solvent extracts of *Aurea Helianthus* flowers contain high phenols and flavonoids content, and exhibited antioxidant activities in vitro, therefore, could be candidates for use as natural antioxidant.

## 1. Introduction

Plants, as valuable natural resources, provide us with food, food additives, flavours, fragrances, colours, as well as pharmaceuticals in medicine [1]. Approximately 50% plant and their formulation products including antibiotics are all drugs available in clinical use in the world, plants and their formulation products constitute more than 35% of the total drugs in use [2]. Presently, antioxidant medicinal plants, including phenols and flavonoids have been showed a wide range of biological activities including anticarcinogenic actions [3]. Medicinal plants with pharmacological properties have been shown to be rich sources of ingredients with significant potential to prevent incurable diseases [4,5]. Many of the plants used traditionally have received little, the medicinal properties of these are undetermined. There is now an urgent need to search for novel and effective chemical compounds that may prevent cancers and other chronic diseases [6]. So, many studies are currently going to develop inhibitors from medicinal plants to prevent or cure chronic inflammatory conditions for minimalside effects [7]. *Aurea Helianthus* is a annual herb in family *Malvaceae*, and has the most edible, medicinal and health care functions in more than 200 kinds of okra plants, and has high commercial value [8].The flowers, stems, leaves and seeds of *Aurea Helianthus* are rich in biologically active substances, which have good anti-oxidation, hypolipidemic, hypoglycemic, anti-tumor, analgesic, antipyretic, anti-inflammatory, immune regulation, anti-aging, Liver protection and other functions[9]. However, studies on phenols and flavonoids content, and antioxidant activities of *Aurea Helianthus* flowers have rarely been reported. Therefore, in order to provide theoretical and experimental basis for development and utilization, our present work is to determine the total phenols and flavonoids content of different polar solvent extracts from *Aurea Helianthus* flowers, and their antioxidant activity.

## 2. Methods

*2.1. Materials*

Ethanol, hexane, chloroform, ethyl acetate and butanol were procured from Junsei (Tokyo, Japan).1, 1-Diphenyl-2-picrylhydrazyl(DPPH), TPTZ (2,4,6-tripyridyl-s-triazine), gallic acid and quercetin were purchased from Sigma-Aldrich Company, Germany. A UV-visible spectrophotometer from Shimadzu, Japan (Model: UV-2450) was used for measuring the absorbance of different polarity crude extracts.

*2.2. Extract sample preparation*

*Aurea Helianthus* flowers were collected from market, 2018. 10. (Pyongyang，Korea). The samples were separated and washed with water. The collected samples were dried at room temperature. The samples were ground to powder. In the conventional reflux extraction method, The samples (20.00 g) was extracted three times with 200 mL of 80 % ethanol aqueous solution at 90 °C, 2 h for each time[10]. Then the extracts were combined, filtered, and concentrated under reduced pressure at 45 °C (7.25 g, yield 36.25%). The obtained ethanol extract(5.00 g) was suspended in water until completely dissolved. Next, the extract was transferred into a separatory funnel and fractionated by hexane, ethyl acetate, chloroform and butanol. The process was repeated 3 times. Every extract were combined and concentrated using a rotary evaporator. The remaining water was also evaporated using a rotary evaporator to give a water extract[11].

*2.3.Determination of total phenols content*

The total phenols content of each extract sample was evaluated using the Foline Ciocalteu reagent (FCR) described by Mc Donald et al [12], with slight modifications. Each extract sample was dissolved in deionized water until 100 μg/mL. The standard curve was obtained using gallic acid (0, 12.5, 25, 50, 100 and 200 μg/mL). Each diluted extract sample or gallic acid (2.0 mL) was added to FCR (0.5 mL, 20 %) and mixed well for 5 minutes. Sodium carbonate (0.5 mL, 10 %) was added to each mixture and all the mixture was kept at room temperature for1 h. The absorbance of all the mixture was measured at 750 nm using a UV-visible spectrophotometer.

*2.4. Determination of total flavonoids content*

The total flavonoids content in each extract sample was investigated using Al $Cl_3$ colorimetry method reported by Zhishen et al [13], with some modifications. Each extract sample was dissolved in methanol to a concentration of 150 μg/mL. The standard curve was prepared by quercetin (0, 12.75, 25.5, 50, 100 and 200 μg/mL). Each diluted extract sample or quercetin (2.5 mL) was added to 6%$NaNO_2$(0.2 mL) and mixed. After 5 min,10% Al Cl3(0.3 ml) was added to each mixture and kept again for 5 min. then, 5% NaOH (1.0 ml) was added to all the mixtures. The absorbance of each mixture was measured at 510 nm using a UV-visible spectrophotometer.

*2.5. Antioxidant activity*

2.5.1 Ferric reducing ability of plasma (FRAP) assay

The FRAP assay of each extract sample was performed following Benzie & Strain (1996)[14], with minor modifications. The FRAP reagents were mixed with 20 ml of 0.3M acetate buffer(p H 3.6), 2.0 ml of 10 mM TPTZ (2,4,6-tripyridyl-s-triazine) in 40 mM HCl, and 2.0 ml of 20 mM FeCl3 6H2O. The extract sample (100μL, 0.5 mg/mL ) was added to 3 mL of FRAP solution and 300 μL of distilled water and then left for 5 min at 37 °C. The absorbance of the extract sample was measured at 593 nm using a UV-visible spectrophotometer. A standard curve was prepared using solutions of FeSO4, Ascorbic acid and BHT were used as positive controls. all experiments were carried out in triplicate, the FRAP values were expressed as mM FeSO4 per gram of extract (mM FS/g extract).

2.5.2. Total antioxidant activity (TAA)

The total antioxidant activity(TAA) of each extract sample was evaluated using the method described by Prieto, Pineda, and Aguilar (1999)[15]. 1 mL of the extract sample(100-500 μg/mL) was mixed with 3 ml of reagent solution (0.6 M sulfuric acid, 28 mM sodium phosphate and 4 mM ammonium molybdate). The mixture was incubated for 90 min at 95 ºC, then the mixture was cooled to room temperature. The absorbance of this mixture was measured at 695 nm using a UV-visible spectrophotometer, indicated antioxidant activity of each extract sample. BHT was used as a positive control, all experiments were carried out in triplicate.

2.5.3. DPPH radical scavenging assay

DPPH radical scavenging assay of each extract sample was carried out by using a modified method ( Gow-Chin Yen and Hui-Yin Chen ,1995)[16]. DPPH solution was prepared by dissolving 10 mg DPPH in 100 mL methanol (about 0.25 mM). 2.0 mL of each extract sample (100-500 μg/mL) was mixed with 2.0 mL of DPPH solution. The mixture was incubated at room temperature in the dark. After 60 minutes, the absorbance was determined at 517 nm and ascorbic acid was used as a positive control. The DPPH radical scavenging ability of each extract sample was calculated using the following equation:

$$\text{DPPH radical scavenging rate}(\%) = [(A_{control} - A_{sample})/A_{control}] \times 100\%$$

where $A_{control}$ is the absorbance of DPPH solution (without extract sample), $A_{sample}$ is the absorbance of extract sample (containing DPPH solution). The antioxidant activity of each extract sample was expressed as $IC_{50}$ (mg ml$^{-1}$), all experiments were carried out in triplicate.

2.5.4. Statistical analyses

All tests were measured in triplicate, and these results were expressed as means ±SD (Standard Deviation). Statistical analysis was performed by SPSS software (version 22.0, USA), and P values < 0.05 was used for statistical significance.

## 3. Results and discussion

*3.1. Extraction yield*

Extraction is the main step for recovering and isolating phytochemicals from plant materials, extraction yield is affected by the chemical nature of phytochemicals, the extraction method used, the solvent used, as well as the presence of interfering substances[17].Each extract sample obtained from ethanol extract (5.00 g) was given to following yields: hexane (0.34 g, yield 6.8%), ethyl acetate (1.12 g, yield 22.4%), chloroform (0.46 g, yield 9.2%), butanol (0.33 g, yield 6.6%) and water(2.62 g, yield 52.4%).

*3.2. Total phenols content*

The total phenols content values of each extract sample, were determined from the standard curve ($r^2$=0.9968) , were expressed as mg gallic acid equivalent per gram of extract (mg GAE/g extract)[12].

As shown in Table 1, the total phenols content of each extract sample followed the order: ethyl acetate (516.21 mg GAE/g) > butanol (246.05 mg GAE/g) > ethanol (225.23 mg GAE/g)> chloroform (212.16 mg GAE/g) > hexane(103.68 mg GAE/g) > water(73.68 mg GAE/g).The highest total phenols content was obtained in ethyl acetate extract , with a trend similar to that of total phenols content obtained from Z. jujuba leaf extract [18].The lowest total phenols content was obtained in water extract, this may be attributable to the content of more nonphenol compounds such as carbohydrate and terpene in water extracts than in other extracts[19].

**Table 1**
Total phenols and flavonoids content in each extract sample of *Aurea Helianthus* flowers

| Extract Sample | Total phenols (mg GAE/g) | Total flavonoids (mg QCE /g) |
|---|---|---|
| Ethanol | 225.23±0.24 | 141.25±0.34 |
| Hexane | 103.68±0.53 | 32.53±0.22 |
| Ethyl acetate | 516.21±0.81 | 328.06±0.65 |
| Chloroform | 212.16±0.36 | 186.28±0.72 |
| Butanol | 246.05±0.62 | 167.32±0.28 |
| Water | 73.68±0.45 | 51.21±0.31 |

All values are presented as mean ± SD of triplicates analysis: GAE: gallic acid, QCE: quercetin.

### 3.3. Total flavonoids content

The total flavonoids content of each extract sample was shown in Table 1, determined from the standard curve ($r^2$=0.9983), was expressed as mg quercetin per gram of extract (mg QCE /g extract) [13]. The total flavonoids content of each extract sample followed the order: ethyl acetate(328.06 mg QCE /g）＞chloroform(186.28 mg QCE /g）＞ethanol (141.25 mg QCE /g）>butanol(167.32 mg QCE /g）> water (51.21mg QCE /g）> hexane(32.53 mg QCE /g). The highest flavonoids content was obtained in ethyl acetate extract and the lowest flavonoids content was obtained in hexane extract.

The correlation between TPC and TFC assay was found to be 0.967. This indicates that flavonoids are the dominating phenolic group in *Aurea Helianthus* [20,21]. Some flavonoids that were isolated from *Aurea Helianthus* have been identified by BAI Yun et al [22].

### 3.4. antioxidant activities

#### 3.4.1 FRAP assay

FRAP assay is often used to evaluate the ability of antioxidant agent to donate electrons, the FRAP values were expressed as mM FeSO4 per gram of extract (mM FS/g extract) [23]. The FRAP value of each extract sample followed the order: ethyl acetate> chloroform > ethanol> hexane > butanol >water (Table 2).

**Table 2**
FRAP value of each extract sample from *Aurea Helianthus* flowers

| Extract Sample | Mean value±SD(mM FS/g) |
|---|---|
| Ethanol | 3.16±0.03[a] |
| Hexane | 2.53±0.05[c] |
| Ethyl acetate | 6.54±0.06[c] |
| Chloroform | 3.68±0.04[a] |
| Butanol | 2.31±0.05[b] |
| Water | 1.62±0.04[d] |
| Ascorbic acid | 12.35±0.08[a] |
| BHT | 2.83±0.05[c] |

All values are presented as means±SD of triplicates analysis, means in the same column with different letters([a-d])were significantly different from one another($p < 0.05$).

FRAP values of ethyl acetate extract was 6.54±0.06 mM FS/g, which was the highest of that of all Extract Samples. FRAP values of water extract was 1.62±0.04 mM FS/g, the lowest of them. FRAP values of all extract samples were correlated to total phenols content (0.912) and flavonoids content(0.866).

#### 3.4.2. TAA

The total antioxidant activity (TAA) of each extract sample was calculated based on the formation of the phosphomolybdenum complex, was measured spectrophotometrically at 695 nm using a UV-visible spectrophotometer [6].

As shown in Fig. 1, The TAA of each extract sample increased with increasing concentration. Ethyl acetate extract showed the highest TAA, wich was significantly different from that of the others (p <0.05). The TAA of each extract sample followed the order: ethyl acetate > chloroform > BHT > ethanol > butanol > hexane > water.

3.4.3. DPPH radical scavenging activity

DPPH radical is a stable free radical with a characteristic absorption at 517 nm (Soares et al, 1997)[24].DPPH radical scavenging activity has been extensively used for screening antioxidant activities of various natural products. The mechanism of the reaction between antioxidant and DPPH radical depends on the structural conformation of the antioxidant(Bondent et al,1997)[25].

As shown in Fig. 2, DPPH radical scavenging activity of each extract sample followed the order: Ascorbic Acid > ethyl acetate> chloroform > butanol > ethanol > hexane > water, also increased with increasing concentration.

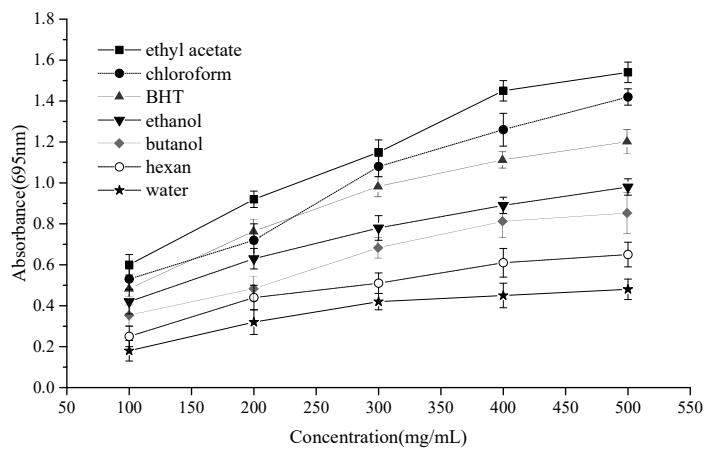

**Fig.1. Total antioxidant activity of each extract sample from *Aurea Helianthus* flowers. All results are means±SD of triplicates analysis. (p < 0.05)**

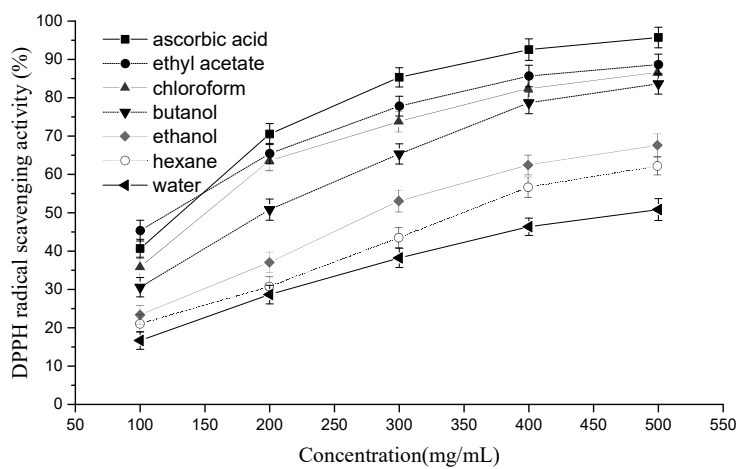

**Fig.2. DPPH radical scavenging activity of each extract sample from *Aurea Helianthus* flowers. All results are means±SD of triplicates analysis. (p < 0.05)**

DPPH radical scavenging activity of Ethyl acetate extract and chloroform extract (except Ascorbic Acid) were significantly different from that of the others (p <0.05).The difference in the DPPH radical scavenging effect of the solvent extracts observed is consistent with research by Bonoli et al.(2004)[26] and Zhao et al. (2006)[27], it was generally found that this antioxidant activity was related to the content of total phenols and total flavonoids of the extracts.

### 3.4.4. IC50

The IC50 of each extract sample is inversely related to its antioxidant capacity, as it expresses the amount of antioxidant required to decrease the DPPH concentration by 50% [28]. The IC50 values in the DPPH radical scavenging activity assay of each extract sample were showed in Table 3.

As shown in Fig. 3, It was found that ethyl acetate extract(except Ascorbic Acid) possesses the strongest DPPH radical activity(IC50 =132.63μg/mL)and water extract possesses the weakest DPPH radical activity(IC50 =492.36μg/mL). Phenols and flavonoids were the main antioxidant components, and their total contents were directly proportional to their antioxidant activity [28]. $IC_{50}$ values of all extract samples from *Aurea Helianthus* flowers were correlated to total phenols content (0.776) and flavonoids content(0.843).

**Table 3**
$IC_{50}$ values of each extract sample from *Aurea Helianthus* flowers

| Extract Sample | $IC_{50}$(μg/mL) |
| --- | --- |
| Ethanol | 283.61±1.34[b] |
| Hexane | 353.62±1.86[a] |
| Ethyl acetate | 132.63±0.82[e] |
| Chloroform | 151.34 ±1.46[d] |
| Butanol | 198.53±1.63[c] |
| Water | 492.36±2.21[f] |
| Ascorbic acid | 123.12±0.64[b] |

All values are presented as means±SD of triplicates analysis, means in the same column with different letters([a-f])were significantly different from one another(p > 0.05).

This result is different from the results of previous studies [29], the differences may be attributed to different solvents used in extraction [30], method and conditions of extraction[31]and so on.

### 4. Conclusion

This study indicated that different polarity extracts of *Aurea Helianthus* flowers contain high phenols and flavonoids content, and exhibited antioxidant activities. In this work, maximum amount of total phenols and flavonoids content were obtained in ethyl acetate and chloroform extracts, also showed significant antioxidant activity. The antioxidant activity was determined using the FRAP assay, TAA assay, DPPH radical scavenging assay. There is a good correlation of total phenols and flavonoids content with antioxidant activity. Therefore, the extracts of *Aurea Helianthus* flowers, as a natural antioxidant, could be used for the treatment of diseases related to oxidative stress.

### Acknowledgments

This work was a supported by the National Key Research Development Program [Grant No. 2016YFD0400600].